\documentstyle[preprint,pre,aps]{revtex}

\begin{document}
\draft
\title{Scaling Relations for Diversity of Languages}
  
\author{ 
M.A.F.Gomes$^1$, 
G.L.Vasconcelos$^1$,
I.J.Tsang$^2$, and I.R.Tsang$^2$ }

\address{
$^1$ Departamento de F\'{\i}sica,
Universidade Federal de Pernambuco,  
50670-901 Recife, PE, Brazil \\
$^2$ Department of Physics,  
University of Antwerp,
Antwerp B-2020 Belgium}
\date{\today} 
\maketitle
\begin{abstract}
The distribution of living languages is investigated and scaling
relations are found for the diversity of languages as a function of
country area and population. These results are compared with data from
Ecology and from computer simulations of fragmentation dynamics where
similar scalings appear. The language size distribution is also
studied and shown to display two scaling regions: (i) one for
the largest (in population) languages and (ii) another one for
intermediate-size languages. It is then argued that these two classes
of languages may have distinct growth dynamics, being distributed on
sets of different fractal dimensions.

\end{abstract}

\section{Introduction}

A great deal of effort has been made to know the Earth's biodiversity
\cite{biodiversity}. In spite of this, only about 1.7 
million of an estimated 13.6 million species have been identified to
date. The diversity of languages, on the other hand, is much better
known: there are 228 countries in the world with a total of
approximately 5000 ethnic groups speaking about 6500 different
languages \cite{sciam,ethnologue}. In this Paper we report a
quantitative analysis of how language diversity increases with country
area and population. A study of the language size distribution is also
presented.

The concept of diversity plays an important r\^ole in an increasing
number of contexts in the scientific literature in connection with
biological problems \cite{bio}, cellular automata \cite{CA}, diffusion
processes \cite{frag1}, ecological \cite{ecology} and evolutionary
\cite{evolution} problems, fractals \cite{fractal}, and
fragmentation phenomena in general \cite{frag2}, including nuclear
fragmentation \cite{nuclear}. Scaling relationships between diversity
and the system size has been reported in a number of studies on
fragmentation \cite{frag1,frag2,frag3,frag4} and Ecology
\cite{ecology,species,model,insetos}. For example, it has now been firmly
established, from both ecological field data \cite{species} and
computer models \cite{ecology,model}, that the number of species or
diversity $D$ in a ecosystem of area $A$ increases with $A$ as a power
law: $D\sim A^z$, where the exponent $z$ varies from 0.1 to 0.45
\cite{species,model}.

Here we report a scaling relationship between language diversity and
area that is akin to the relation above observed in Ecology. A scaling
relation has also been found between language diversity and
population.  The diversity distribution, meaning the number
of countries with a given diversity, displays composite power-laws,
and an argument is presented to account for the existence of
these two distinct scaling regimes. We also study the language size
distribution and show that it has two scaling regions corresponding,
respectively, to (i) the largest (in population) languages and (ii)
intermediate-sized languages. It is then argued that the existence of
distinct scaling behaviors for these two classes of languages imply
that they have distinct growth dynamics, leading to different patterns
of space occupation, such as different fractal dimensions.

\section{Results and Discussion}

Our study was based on the thirteenth edition of the Ethnologue
\cite{ethnologue}, published in 1996, which lists more than 6700
languages spoken in 228 countries.  We divided the countries in 12
groups (bins) according to area and then calculated the average
diversity $D$ of living languages in each bin.  In Fig.\ 1 we plot our
results for language diversity as a function of area.  As one sees in
this figure, the data points are well fitted by a power law:
\begin{equation}
D \sim A^{z}, \label{eq:1}
\end{equation}
with $z = 0.41 \pm 0.03$, an exponent close to the largest
values found in Ecology \cite{ecology}. It should be emphasized that
the power law shown in Fig.\ 1 extends over almost six decades, the
only deviation occurring for countries with area smaller than 30
km$^2$. The density of living languages, $\rho_D=D/A$, thus scales as
$\rho_D\sim A^{-0.59}$, implying that larger areas have
proportionately less diversity of languages.

We have also studied how language diversity varies with population. In
Fig.\ 2 we show the dependence of the average language
diversity $D$ as a function of the average population $N$ within each
area bin. In this case we find a power-law of the type:
\begin{equation}
D \sim N^\nu, \label{eq:2}
\end{equation}
where from a best fit we obtain $\nu=0.50\pm0.04$. It is interesting
to note that similar scaling (with $\nu=1/2$) between diversity and
population has been found in computer simulations and experiments on
fragmentation dynamics \cite{frag1,frag2,frag3,frag4} as well as in insect
populations \cite{insetos,us}. Figure 2 also shows that on average a group of
about 15,000 people is needed to maintain one single language
alive. This might be of relevance {\it vis-\`a-vis} the potential danger of
extinction of several languages \cite{extinction} whose number of
native speakers are presently well below this threshold.

From Eqs.\ (1) and (2) it follows that the (average) population
grows with the (average) country area as a power law:
\begin{equation}
N \sim A^{z/\nu}. \label{eq:3}
\end{equation}
From the values for $z$ and $\nu$ above we thus obtain that $N \sim
A^{0.82}$.  (This result could also have been obtained directly from
the data on country areas and populations without referring to the
language distribution). In other words, the (average) population
density on Earth, $\rho_N=N/A$ is not constant but rather decreases
with area as $\rho_N\sim A^{-0.18}$. Thus countries with large areas
are proportionately less populated than smaller ones, as is widely
known. Moreover, from Eq.\ (\ref{eq:3}) and the fact that $A\sim L^2$,
where $L$ is a linear length scale, it then follows that
\begin{equation}
N\sim L^d, \label{eq:4}
\end{equation}
where $d=2z/\nu=1.64$, thus indicating that the human population is
distributed over the surface of the earth on a fractal set of
dimension $d=1.64$. Note also that from Eq.\ (1)
and the fact that $A\sim L^2$ it follows that language diversity
scales with linear size as
\begin{equation}
D \sim L^\delta, \label{eq:6}
\end{equation}
where $\delta=2z=0.82$, meaning that living
languages are distributed on a set of dimension close to unity.
 
Another interesting pattern concerns the distribution of language
diversity among the various countries. We show in Fig.\ 3 the
cumulative diversity distribution, ${\cal N}(>$$D)$, corresponding to
the number of countries with a language diversity greater than $D$. We
see from this figure that ${\cal N}(>\!D)$ displays composite
power-laws: ${\cal N}(>\!D)\sim D^{- B}$, with $B=0.6$ for $6 < D <
60$ and $B=1.1$ for $60< D< 700$ (each power-law in this case extends over one
decade or more.)  Now, why should the diversity distribution have two scaling
regimes with ${\cal N}(>\!D)$ decreasing faster for larger $D$? A
possible answer is that it is difficult in general to preserve the
unity of large countries with great language (and hence ethnic)
diversity, since they will tend to break up into smaller ones.  This
process could thus account, at least in part, for the fact that the
cumulative diversity distribution ${\cal N}(>\!D)$ crosses over to a
faster decay as $D$ increases.

Scaling relations (\ref{eq:1}) and (\ref{eq:2}) above were obtained
averaging the language diversity and the population over countries of
similar area. To obtain a better estimate of the exponents $z$ and
$\nu$ one should ideally count the language diversity and total
population contained in concentric regions, say, circles, of increasing
size, as is costumary in statistical physics.  Unfortunately, this
procedure would be quite cumbersome here, if possible at all, and so
we had to resort to the data reported in the Ethnologue
\cite{ethnologue} for individual countries.  We believe, however, that
the persistence of scaling behavior over several decades in Figs.\ 1
and 2 is an indication that our estimates are statistically reliable.

We have also studied the language size distribution---a quantity that
does not directly depend on geopolitical boundaries. In
Fig.\ 4 we show the cumulative size distribution, $n(>\!N)$,
corresponding to the number of languages with a population greater
than $N$. We see from this figure that $n(>\!N)$ displays composite
power-laws:
\begin{equation}
n(>\!N) \sim N^{-\tau}, \label{eq:7}
\end{equation}
with $\tau=0.5$ for $5\times10^4 < N < 6\times10^6$, and $\tau=1.0$
for $2\times 10^7 < N < 1\times10^9$.  Note that each of these power
laws is valid for about two decades.

The fact that the exponents $\tau$ for the largest and
intermediate-sized languages differ might be seen as an evidence that
these two classes of languages possess distinct growth dynamics,
leading to different patterns in the occupation of space. To see
this, we first introduce the fractal dimension $\cal D$ defined by
\cite{fractal}:
\begin{equation} 
n(>\!L) \sim L^{-\cal D}, \label{eq:8} 
\end{equation} 
where $n(>\!L$) is the number of languages that occupy a region of
linear size greater than $L$.  From Eqs. (\ref{eq:4}), (\ref{eq:7}), and
(\ref{eq:8}) one then immediately finds :
\begin{equation} 
{\cal D} = d \tau. \label{eq:5}
\end{equation} 
Thus, the languages with largest populations, for which $\tau=1$, may
be regarded as `space filling' in the sense that ${\cal D}=d$, i.e.,
they are distributed on a subset of dimension equal to the dimension
of the set on which the entire population is distributed.  On the
other hand, languages with smaller populations ($\tau=0.5$) are more
`tenuously' distributed on the surface of Earth, since they occupy a
subset of dimension (${\cal D}=d/2=0.82$) considerably less than $d$.
Note also that in this case ${\cal D}=\delta$ [see Eq.\
({\ref{eq:6})], thus showing that the dominant contribution to language
diversity comes, as expected, from languages with small to
intermediate-sized populations.  The results above conform with the obvious
fact that languages with large populations tend to be more widely
spread, whereas languages with smaller populations are in general
restricted to small areas (most of the languages in the region with
$\tau=0.5$ in Fig.\ 4 are indeed confined within a single country). It
is also interesting to notice that the exponent $\tau=1$ for languages
with large populations is similar to what is usually found in
classical critical phenomena
\cite{tau}.  Of course, a more detailed model for population dynamics
is required if one wishes to explain, in a more quantitative fashion,
the interesting features revealed by
the present analysis.  We are currently working on this direction.

\section{Conclusions}

We have presented a quantitative analysis of the diversity of human
languages.  We have studied how language diversity increases with area
and population and found scaling relations in both cases. The language
size distribution was also analyzed and shown to display two distinct
power laws: (i) one with the exponent $\tau=1$ for the top 50
languages by population and (ii) another one with $\tau=0.5$ for
languages with population between fifty thousand and six million.  The
corresponding fractal dimension $\cal D$ for these two classes of
languages was obtained, and it was found that the largest languages
are `space filling' (${\cal D}=d=1.64$) with respect to the set
available for the entire population, whereas intermediate-sized
languages are more thinly distributed (${\cal D}=d/2=0.82$) but give
the main contribution to language diversity.

This work was partially supported by Conselho Nacional de
Desenvolvimento Cient\'{\i}fico e Tecnol\'ogico and Financiadora de
Estudos e Projetos (Brazilian Agencies).

\begin{figure}
\caption{Average diversity of languages as a function of area. The straight 
line is a best fit whose slope gives the exponent $z=0.41\pm0.03$.}
\end{figure}

\begin{figure}
\caption{Average diversity of languages as a function of average 
population within a bin area. The solid line is a best fit with slope
$\nu=0.50\pm0.04$.}
\end{figure}

\begin{figure}
\caption{Number of countries with a language diversity greater than $D$ 
as a function of $D$. The straight lines give the exponents $B=0.6$
for $6 < D < 60$ and $B=1.1$ for $60< D< 700$.}
\end{figure}

\begin{figure}
\caption{Number of languages with population greater than 
$N$ as a function of $N$. The straight lines give the exponents  $\tau=0.5$ for
$5\times10^4 < N < 6\times10^6$ and $\tau=1.0$ for $2\times 10^7 < N 
< 1\times 10^9$.}

\end{figure}


\begin{thebibliography}{}

\bibitem{biodiversity} S. Blackmore, Science {\bf 274} (1996) 63.

\bibitem{sciam} R. Doyle, Sci. Am. {\bf 279} (1998) 19.

\bibitem{ethnologue} http://www.sil.org/ethnologue.

\bibitem{bio} D. W. Thompson, On Growth and Form, Cambridge
University Press, Cambridge, 1971.

\bibitem{CA} 
S. Wolfram, Theory and Applications of Cellular Automata,
World Scientific, Singapore, 1986.

\bibitem{frag1} K. R. Coutinho, M. D. Coutinho-Filho, M. A. F. Gomes, 
A. N. Nemirovsky, Phys. Rev. Lett. {\bf 72} (1994) 3745.

\bibitem{ecology} I. Hanski and M. Gyllenberg, {\it Science} {\bf 275} (1997) 397.

\bibitem{evolution} D. M. Raup, S. J. Gould, T. J. Schopf, D. S. Simberloff, J. Geology {\bf 81} (1973) 525.

\bibitem{fractal}
B. B. Mandelbrot, The Fractal Geometry of Nature, Freeman, New York,
1983.

\bibitem{frag2}
K. Coutinho, S. K. Adhikari, M. A. F. Gomes, J. Appl. Phys. {\bf 74}
(1993) 7577.

\bibitem{nuclear} C. Lewenkopf, J. Dreute, A. A.-Magd, J. Aichelin, W. 
Heinrich, J. H\"ufner, G. Rusch, B. Wiegel, Phys. Rev. C {\bf 44} (1991) 1065.

\bibitem{frag3}K. Coutinho, M. A. F. Gomes, S. K. Adhikari, 
Europhys. Lett. {\bf 18} (1992) 119.

\bibitem{frag4} V. P. Brito, M. A. F. Gomes, F. A. O. Souza, S. K. Adhikari, 
Physica A {\bf 259} (1998) 227.

\bibitem{species} M. L. Rosenzweig, Species Diversity in Space and Time, 
Cambridge Univ. Press, Cambridge, 1995.  

\bibitem{model} J. D. Pelletier, Phys. Rev. Lett. {\bf 82} (1999) 1983.

\bibitem{insetos}
E. Siemann, D. Tilman, J. Haarstad, Nature {\bf 380} (1996) 704.

\bibitem{us} M. A. F. Gomes, G. L. Vasconcelos, S. K. Adhikari, I. J. Tsang, 
I. R. Tsang, Proceedings of the XV SITGES Euroconference--Statistical
Mechanics of Biocomplexity, Barcelona, June 1998, pp. 33-35.

\bibitem{extinction} J. E. Grimes, Oceanic Linguistics {\bf 34} (1995) 1.

\bibitem{tau} M. E. Fischer, Proc. Int. Sch. Phys. Enrico Fermi, Course LI, 
Critical Phenomena, M. S. Green (Ed.), Academic Press, New York, 1971.
  
\end{thebibliography}
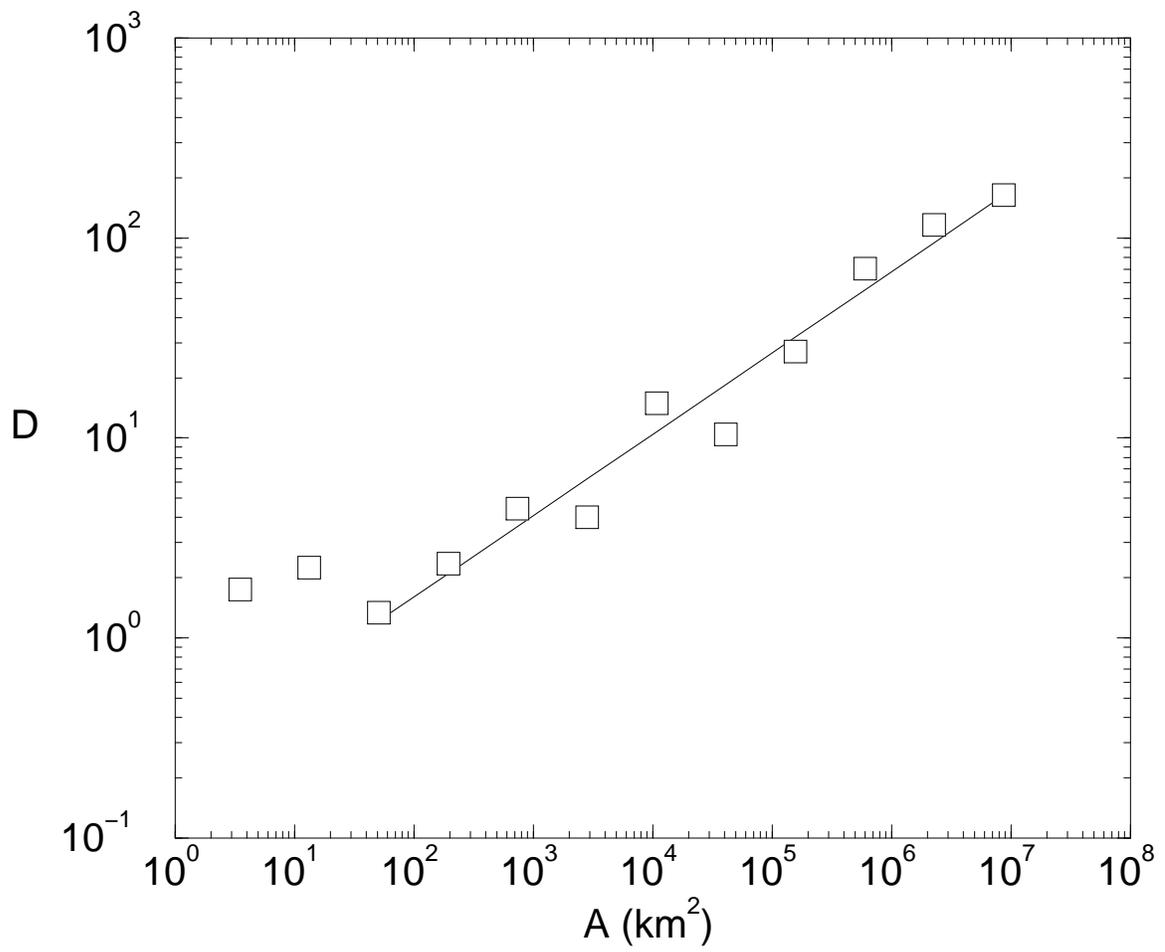
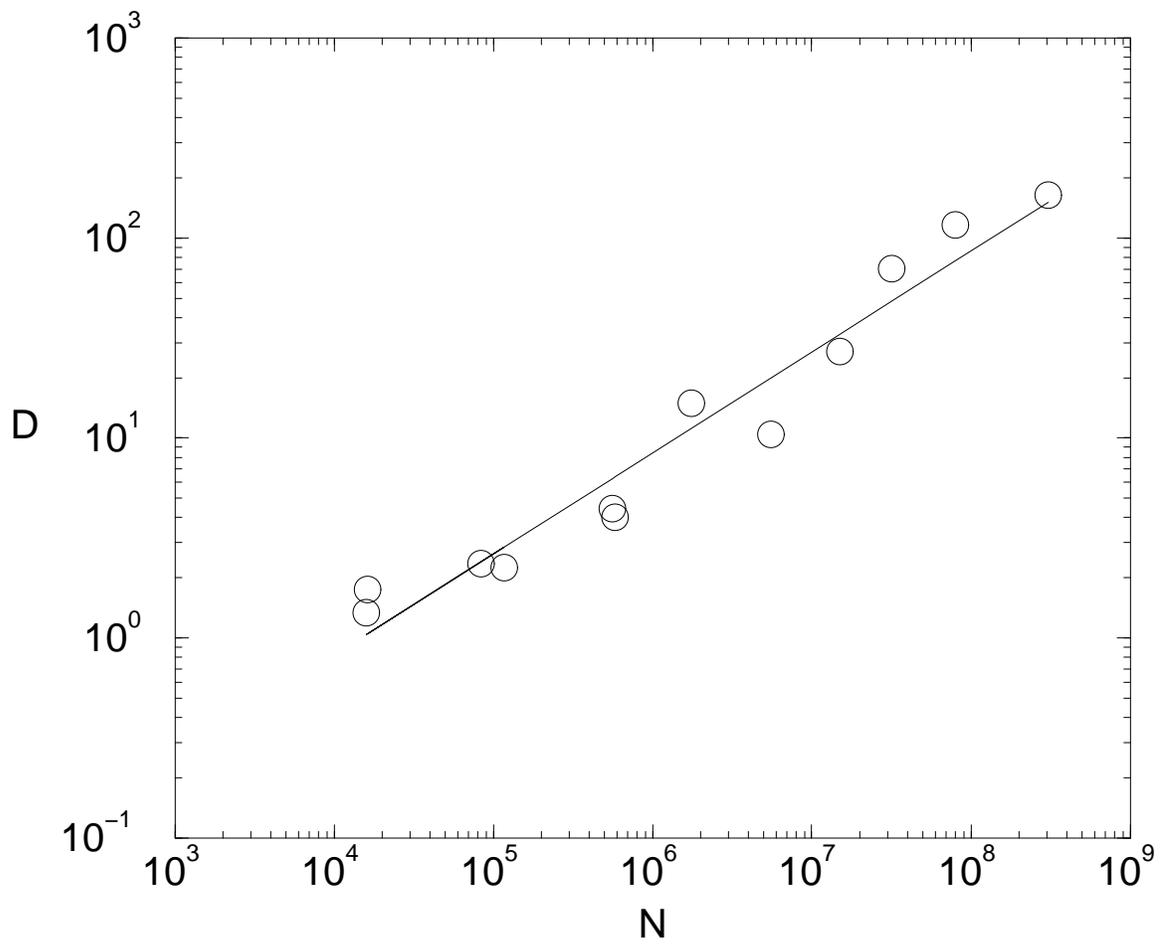
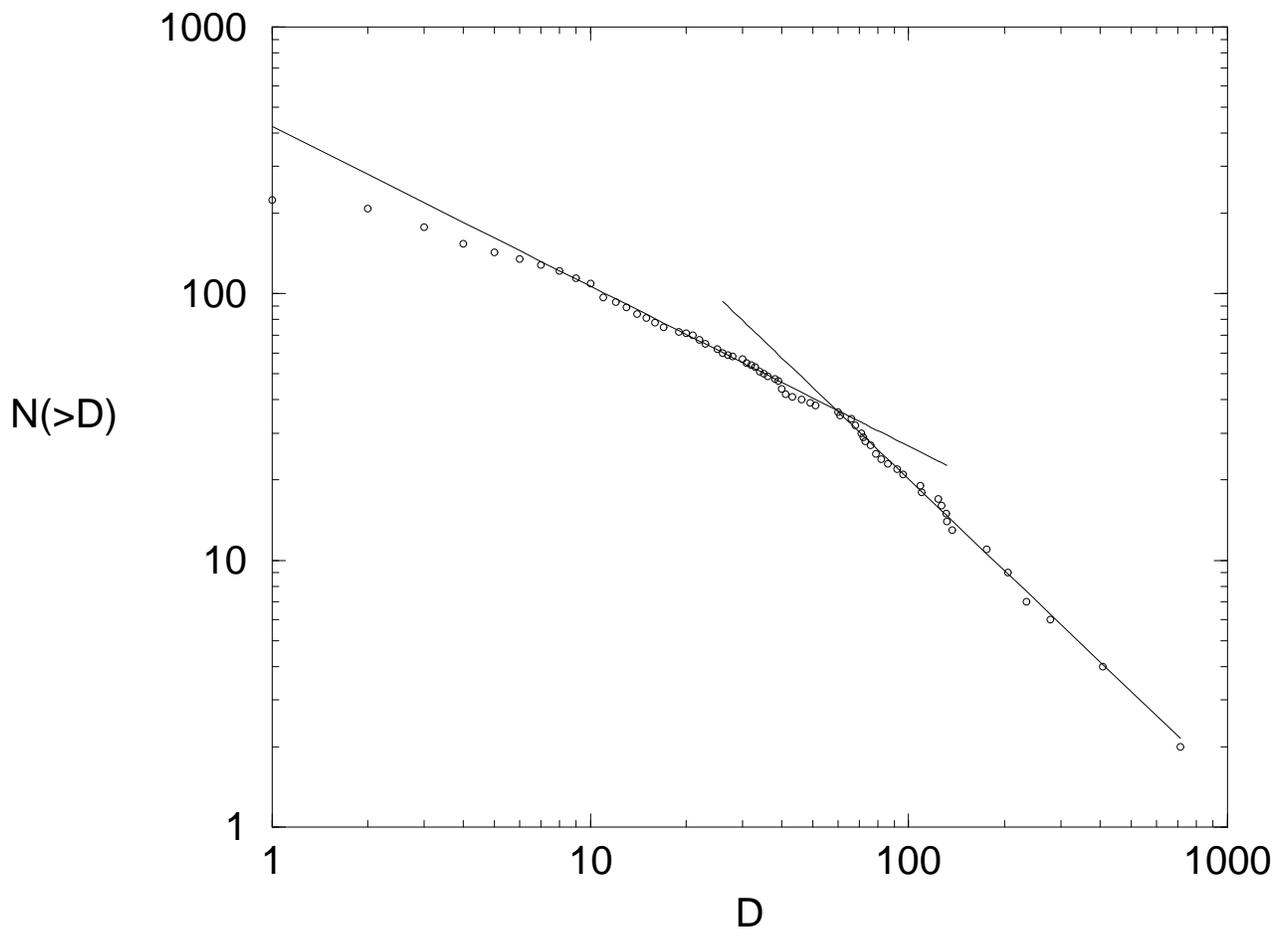
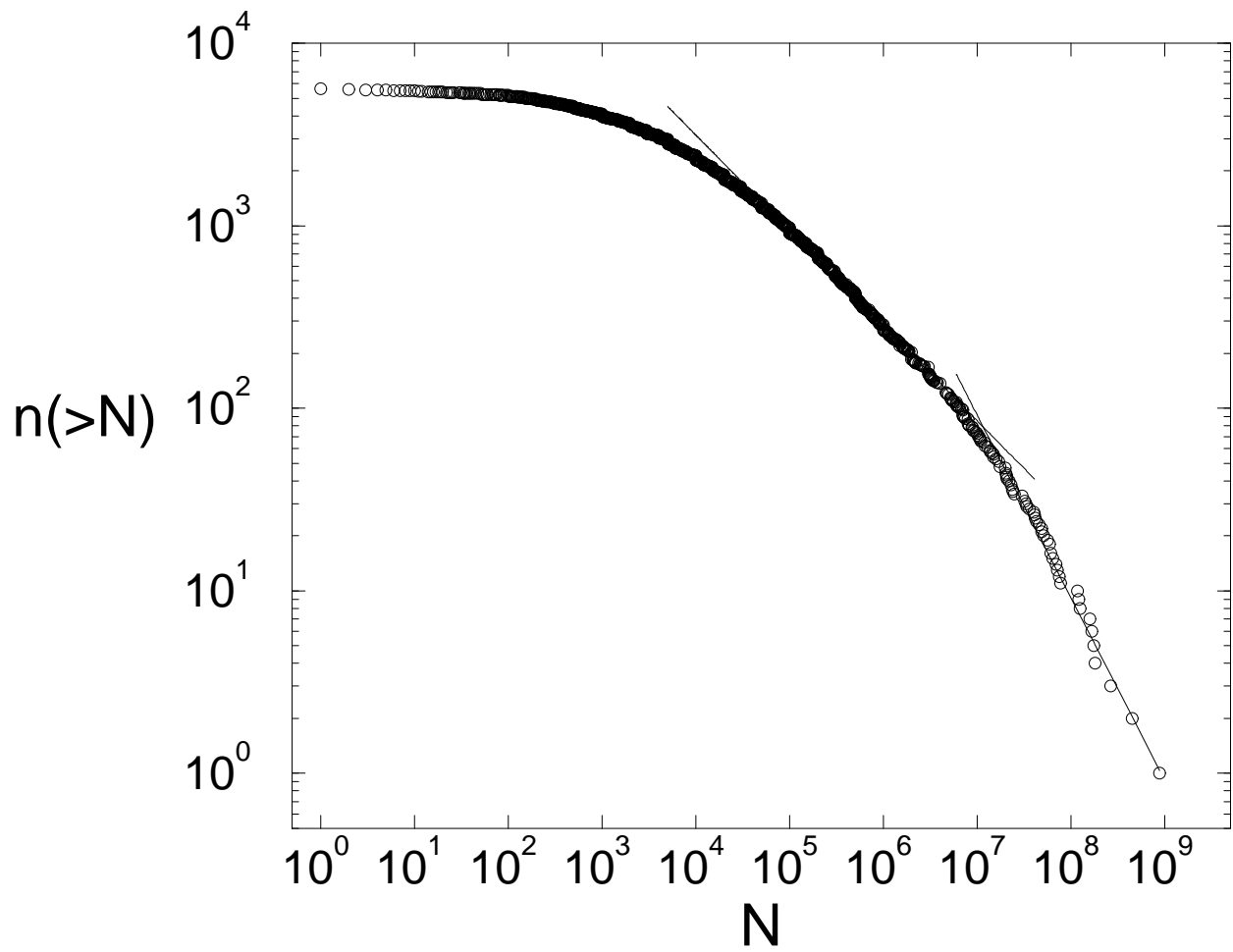
\end{document}